\documentclass[aps,reprint,showpacs,nofootinbib,superscriptaddress]{revtex4-1}

\newcommand\NEG{\vspace*{-1ex}}

\usepackage{epsfig,amssymb,amsmath,latexsym,hyperref}
\usepackage{color}
\definecolor{nicered}{rgb}{0.9,0.1,0.1}

\newcommand{\be}{\begin{equation}}
\newcommand{\ee}{\end{equation}}
\newcommand{\bea}{\begin{eqnarray}}
\newcommand{\eea}{\end{eqnarray}}
\newcommand{\no}{\noindent}

\newcommand\B{{\mathcal B}}

\renewcommand\P{\ensuremath{\mathcal P}}
\newcommand\C{\ensuremath{\mathcal  C}}

\renewcommand\l{\lambda}

\renewcommand\a{\alpha}
\renewcommand\b{\beta}
\newcommand\e{\epsilon}
\newcommand\m{\mu}

\newcommand\g{\gamma}

\newcommand\tr{\text{tr}}

\newcommand\MeV{\text{MeV}}
\newcommand\GeV{\text{GeV}}
\newcommand\TeV{\text{TeV}}
\renewcommand\t{\ensuremath{\theta}}
\renewcommand\l{\ensuremath{\lambda}}

\newcommand\ba{\begin{array}}
\newcommand\ea{\end{array}}

\makeatletter
\def\frontmatter@title@above{\null\addvspace{1\baselineskip}}
\def\frontmatter@title@below{\addvspace{1\baselineskip}}
\def  \frontmatter@authorbelow{\show\pippo\addvspace{10\baselineskip}}
\def\frontmatter@preabstractspace{2\baselineskip}
\def\frontmatter@postabstractspace{2\baselineskip}
\makeatother

\begin{document}
\baselineskip=1.06\baselineskip
\parskip.2ex

\title{New physics in $\e'$ from chromomagnetic contributions\\
  and limits on Left-Right symmetry}

\author{S. Bertolini}
\affiliation{INFN, Sezione di Trieste, 
Italy}
\author{J. O. Eeg}
\affiliation{Department of Physics, University of Oslo, 
Norway}
\author{A. Maiezza}
\affiliation{Dipartimento di Fisica, Universit\`a di L'Aquila, Italy}
\affiliation{INFN, Laboratori Nazionali del Gran Sasso, Assergi, Italy}
\author{F. Nesti}
\affiliation{Dipartimento di Fisica, Universit\`a di L'Aquila, Italy}

\begin{abstract}
  \no New physics in the chromomagnetic flavor changing transition $s\to d g^*$ can avoid the strong
  GIM suppression of the Standard Model and lead to large contributions to CP-violating observables,
  in particular to the $\epsilon'$ parameter, that we address here.  We discuss the case of the
  Left-Right symmetric models, where this contribution implies bounds on the phases of the
  right-handed quark mixing matrix, or in generic models with large phases a strong bound on the
  Left-Right symmetry scale.  To the leading order, a numeric formula for $\e'$ as a function of the
  short-distance coefficients for a wide class of models of new physics is given.
\end{abstract}



\maketitle 

\newpage



\no Flavor-changing processes still offer one of the best means for spotting signs of physics beyond
the Standard Model (SM). The $K$ decays into pions are among the best studied channels both
experimentally and theoretically, and despite the hadronic uncertainties in the theoretical
predictions, can serve as a great tool for probing new physics. The reason is that for a number of
observables, in particular the CP violating ones, the SM contribution is extremely small, mainly due
to the GIM mechanism~\cite{GIM} and the fact that the CP violating phase in the SM is suppressed by
the smallness of the quark mixing angles.  In turn, in the SM the GIM mechanism is intimately
related to the chiral nature of the weak interactions. As a result, probes of processes involving
the GIM mechanism are well suited to test for nonchiral interactions.

This is paradigmatic in one popular extension of the SM, Left-Right (LR) symmetry~\cite{lrmodel},
which altogether gives a framework for restoration of parity in fundamental interactions, nonzero
neutrino masses, as well as violation of lepton number and flavor~\cite{LRLNV} both at the reach of
the coming round of experiments, fitting especially well with the scenario of TeV scale
LR-symmetry~\cite{noi,Tello:2010am, Nemevsek:2011aa}.  The related direct searches at LHC, with
important signatures through the new interactions and same-sign dileptons~\cite{KS}, can explore
this possibility up to $\sim$6\,\TeV~\cite{Ferrari:2000sp} and are already beginning to probe this
interesting region~\cite{Nemevsek:2011hz,CMS,Degenhardt:2011aw}. It is then important to assess the
bounds on the model from existing phenomena.

In the LR models, based the $SU(2)_L\times SU(2)_R\times U(1)_{B-L}$ gauge group, modifications of
GIM are mainly due to the new right-handed gauge boson $W_R$ and to its mixing with the standard
weak gauge boson $W_L$.  
%
Bounds on the scale of the new right-handed gauge interaction were already addressed since the early
days, a notorious example being the $\Delta M_K$ box diagram~\cite{soni, Mohapatra:1983ae} where the
GIM enhancement adds to a chiral enhancement of the matrix elements, and still leads today to the
strongest bound on the scale of LR-symmetry, $M_{W_R}\gtrsim
2.5\text{--}3\,\TeV$~\cite{noi}. Similar effects hold for the CP-violation parameter
$\e$~\cite{Ecker:1985vv}.  The bottom line is that the interplay of nonchiral interactions with the
hierarchy of quark masses and mixings can lead to dramatic effects in loop diagrams.  This is
especially true in the phenomenology of strange mesons, and in particular for the direct
CP-violation parameter $\epsilon'$ and the chromomagnetic loop, that we address.

The gluonic penguin operators have been traditionally associated to $\e'/\e$, because they pointed
immediately to a possibly sizable effect.  However, in the SM a partial cancellation between the
dominant gluonic and electroweak penguins translates into a large theoretical uncertainty, linked to
hadronic matrix elements. For a review on the evaluation of $\e'/\e$ and additional literature we
refer to~\cite{eegbertfabbreview,Buras:2003zz,Pich:2004ee}.  In any case, $\e'$ is naturally tiny in
the SM, and can serve as an efficient tool for constraining new physics.

In this work, we address the contributions of nonchiral interactions in the chromomagnetic operator,
and its effect on $\epsilon'$.  In the analysis we first give a parametrization of the effects of
new physics in $\e'$ which is applicable to a wide class of models with nonchiral interactions.  In
the context of Left-Right symmetry, as is known current-current operators mediated through the
left-right gauge boson mixing gives large contributions to the $K\to \pi\pi$ amplitude. This issue
was studied in detail in~\cite{moha,Chen:2008kt,noi,Blanke:2011ry}, together with the other flavor
constraints on the model.  However, the effect of the chromomagnetic operator was not considered.  We
study its impact due to the effective $K\to\pi\pi$ transition whose hadronic matrix element
computed within the Chiral Quark Model ($\chi$QM)~\cite{chiralQM}.

The $\chi$QM provides an interpolation between short-distance QCD and its effective
description in terms of the octet of Goldstone mesons, below the scale of chiral symmetry breaking
(for a recent discussion see~\cite{deRafael:2011ga}). The chiral lagrangian coefficients are
determined order by order in the momentum expansion by integration of the constituent quarks and
depend on three non perturbative parameters: the constituent quark mass and the quark and gluon
condensates. Via a fit of the $\Delta I=1/2$ rule in $K\to \pi \pi$ decays, the authors of
ref.~\cite{eegbertfabbreview} obtained a non trivial phenomenological determination of these three
parameters that allowed for a correlated calculation of $\e'/\e$ and of the $\Delta S=2$ bag
parameter $B_K$ within the $\chi$QM approach, at next-to-leading order (NLO) in the chiral
expansion~\cite{antonelli,extraberto,Bertolini:1997ir}. We will use these values of the parameters
in our analysis.

For the LR model, we shall see that only the chromomagnetic operator plays a dominant role in $\e'$,
once other existing constraints from $K$ and $B$ physics are considered, and implies a bound on the
free phases involved, in the hypothesis of TeV scale LR-symmetry.  For other new-physics models,
constraints from $\e'$ via the chromomagnetic operator were studied
in~\cite{colangelo,Davidson:2007si, Gedalia:2009ws}.

The paper is organized as follows: In section \ref{sec:op} we describe the effective operators
involved when nonchiral new physics is present, including the dipole ones. We also review the short
distance coefficients in the case of the LR theory.  In section~\ref{sec:running}, by running with
the mixed anomalous dimensions, we compute the Wilson coefficients at 0.8\,\GeV, which is our chosen
scale for matching with chiral perturbation theory.  In section~\ref{sec:bosonization} we describe
(and update) the bosonization of the chromomagnetic operator.  This enables us to make contact with
the $K\to\pi\pi$ amplitude and with $\e'$ in section~\ref{sec:results} in general and in
section~\ref{sec:LR} for the LR model.  In section~\ref{sec:conclusions} we draw our conclusions.

\NEG
\NEG

\section{New Physics}
\NEG

\label{sec:op}

\no The effective lagrangian for flavor changing can be written in the form $L_{\Delta S=1} =-
(G_F/\sqrt{2}) \sum_i C_iQ_i+h.c.$, where $Q_i$ are the relevant operators and $C_i$ the
corresponding coefficients (and $G_F$ the Fermi constant). In the Standard Model the $\Delta S=1$
processes are usually described by a (over)complete set of
operators~\cite{standardoperators,standardoperators2}.  They involve tree-level operators as well as
QED and QCD penguins. In models with both chiralities such as the Left-Right model, the standard set
of operators has to be extended.  In the case of $\Delta S=1$ discussed here the complete set at low
energy involves 28 operators,
\begin{equation}
\small
\label{eq:operators}
\begin{array}{ll}
Q_1^{LL} = (\bar{s}_\alpha u_\beta)_L (\bar{u}_\beta d_\alpha)_L & Q_1^{RR} = (\bar{s}_\alpha u_\beta)_R (\bar{u}_\beta d_\alpha)_R \\[0.2cm]
Q_2^{LL} = (\bar{s} u)_L (\bar{u} d)_L & Q_2^{RR} = (\bar{s} u)_R (\bar{u} d)_R \\[0.2cm]
Q_3      = (\bar{s} d)_L (\bar{q} q)_L &Q'_3      = (\bar{s} d)_R (\bar{q} q)_R \\[0.2cm]
Q_4      = (\bar{s}_\alpha d_\beta)_L (\bar{q}_\beta q_\alpha)_L&Q'_4      = (\bar{s}_\alpha d_\beta)_R (\bar{q}_\beta q_\alpha)_R \\[0.2cm]
Q_9      = \frac{3}{2}(\bar{s} d)_L e_q(\bar{q} q)_L&Q'_9      = \frac{3}{2}(\bar{s} d)_R e_q(\bar{q} q)_R \\[0.2cm]
Q_{10}      = \frac{3}{2}(\bar{s}_\alpha d_\beta)_L e_q(\bar{q}_\beta q_\alpha)_L&Q'_{10}      = \frac{3}{2}(\bar{s}_\alpha d_\beta)_R e_q(\bar{q}_\beta q_\alpha)_R \\[0.5cm]
Q_1^{RL} = (\bar{s}_\alpha u_\beta)_R (\bar{u}_\beta d_\alpha)_L & Q_1^{LR} = (\bar{s}_\alpha u_\beta)_L (\bar{u}_\beta d_\alpha)_R \\[0.2cm]
Q_2^{RL} = (\bar{s} u)_R (\bar{u} d)_L &Q_2^{LR} = (\bar{s} u)_L (\bar{u} d)_R \\[0.2cm]
Q_5      = (\bar{s} d)_L (\bar{q} q)_R&Q'_5      = (\bar{s} d)_R (\bar{q} q)_L \\[0.2cm]
Q_6      = (\bar{s}_\alpha d_\beta)_L (\bar{q}_\beta q_\alpha)_R&Q'_6      = (\bar{s}_\alpha d_\beta)_R (\bar{q}_\beta q_\alpha)_L \\[0.2cm]
Q_7      = \frac{3}{2}(\bar{s} d)_L e_q(\bar{q} q)_R&Q'_7      = \frac{3}{2}(\bar{s} d)_R e_q(\bar{q} q)_L \\[0.2cm]
Q_{8}      = \frac{3}{2}(\bar{s}_\alpha d_\beta)_L e_q(\bar{q}_\beta q_\alpha)_R&Q'_{8}      = \frac{3}{2}(\bar{s}_\alpha d_\beta)_R e_q(\bar{q}_\beta q_\alpha)_L \\[0.5cm]
Q_{g}^{L}  = \frac{g_sm_s}{8\pi^2}\bar{s}\sigma_{\mu\nu}t^a G_a^{\mu\nu}L d& Q_{g}^{R}  = \frac{g_sm_s}{8\pi^2}\bar{s}\sigma_{\mu\nu}t^a G_a^{\mu\nu}R d \\[0.2cm]
Q_{\g}^{L}  = \frac{em_s}{8\pi^2}\bar{s}\sigma_{\mu\nu} F_a^{\mu\nu}L d & Q_{\g}^{R}  = \frac{em_s}{8\pi^2}\bar{s}\sigma_{\mu\nu} F_a^{\mu\nu}R d\,.
\end{array}
\end{equation}%
The notation is $(\bar{q}q)_{L,R}=\bar q \gamma_\mu (1\mp \gamma_5)q$, $L,R=(1\mp\gamma_5)/2$, and
the summation on $q=u,d,s$ is implicit.  $Q_{1,2}^{LL}$ are the SM operators usually denoted as
$Q_{1,2}$. The dipole operators $Q_{g,\gamma}$ are normalized with $m_s$, for an easy comparison
with existing calculations, and for keeping unaltered the anomalous dimension.  §It is known that
some of the operators above are accompanied by an enhancement due to the different chiral structure,
either in the short distance coefficient, in the running, or in the matrix element.  Such situation
occurs for the gluo-magnetic operators $Q_g^{L,R}$, as we describe below.

At leading order the operators generated by the SM and the LR short distance physics are:
$Q_2^{AB}$, $Q_4$, $Q'_4$, $Q_6$, $Q'_6$, $Q_7$, $Q'_7$, $Q_9$, $Q'_9$, $Q^{A}_{g}$, $Q^{A}_{\g}$,
with $A,B=L,R$.  Their coefficients are: the current-current ones
\be
\begin{split}
C_2^{LL} = \lambda_u^{LL}\,,\qquad&
C_2^{LR} = \zeta^* \lambda_u^{LR}\,,\\[.7ex]
C_2^{RR} = \beta \lambda_u^{RR}\,,\;\quad&
C_2^{RL} = \zeta \,\lambda_u^{RL}\,;
\end{split}
\label{eq:C2AB}
\ee
the penguin ones
\be
\begin{split}
&C_4=C_6  = \frac{\alpha_s}{4\pi} \Sigma_i \lambda_i^{LL} F_1^{LL}(x_i)\\[.7ex]
&C'_4=C'_6  = \frac{\alpha_s}{4\pi} \beta\, \Sigma_i \lambda_i^{RR} F_1^{RR}(\beta x_i)\\[.7ex]
&C_7=C_9  = \frac{\alpha e_u}{4\pi} \Sigma_i \lambda_i^{LL} E_1^{LL}(x_i)\\[.7ex]
&C'_7=C'_9  = \frac{\alpha e_u}{4\pi} \beta\, \Sigma_i \lambda_i^{RR} E_1^{RR}(\beta x_i)\,;
\end{split}
\label{eq:C4C9}
\ee
and the dipole ones
{\small
\arraycolsep=.2em
\be
\begin{split}
\!\!\!\!&m_sC_{g}^{L}\! = \Sigma_i\Big[ m_s \lambda_i^{LL} F^{LL}_2 +\!\zeta  m_i \lambda_i^{RL} F^{LR}_2+\!m_d\beta \lambda_i^{RR} F^{RR}_2\Big]\!\! \\[1ex]
\!\!\!\!&m_sC_{g}^{R} \!= \Sigma_i\Big[ m_d \lambda_i^{LL} F^{LL}_2 + \!\zeta^*  m_i \lambda_i^{LR} F^{LR}_2+\!m_s\beta \lambda_i^{RR} F^{RR}_2\Big]\!\!\! \\[1ex]
\!\!\!\!&m_sC_{\g}^{L} \!=\Sigma_i\Big[ m_s\lambda_i^{LL} E^{LL}_2 +\!\zeta  m_i \lambda_i^{RL} E^{LR}_2+\!m_d\beta \lambda_i^{RR} E^{RR}_2\Big]\!\!\! \\[1ex]
\!\!\!\!&m_sC_{\g}^{R} \!= \Sigma_i\Big[ m_d \lambda_i^{LL} E^{LL}_2 +\!\zeta^*  m_i \lambda_i^{LR} E^{LR}_2+\!m_s\beta \lambda_i^{RR} E^{RR}_2\Big]\!\rlap{\,\,.}
\end{split}
\label{eq:Cg12}
\ee}%
In the above, $e_u=2/3$ is the $u$-quark charge, $x_i=m_i^2/m_{W_L}^2$ with $i=u,c,t$, and
$F^{AB}_{(1,2)}$ and $E^{AB}_2$ are the loop functions, given in appendix~\ref{app:loop}. Then,
$\beta = M_{W_L}^2/M_{W_R}^2$ is the ratio of the electroweak to the LR scale and $\zeta$ is the
$W_L$-$W_R$ mixing. Note that in the hypothesis of LR symmetry at TeV scale, $\b\sim 10^{-3}$. Also
$\zeta$ is of order $\b$ or less; for instance in the minimal LR models it is $\zeta\simeq -\beta
e^{i\alpha}\frac{2x}{1+x^2}$, with $x<1$ the (modulus of the) ratio of the two VEVs of the Higgs
bi-doublet, and $\alpha$ its phase. We will consider below the specific case of minimal LR models,
referring to~\cite{noi,Senjanovic:1978ev} for definitions and details.  Finally
$\lambda_i^{AB}=V_{is}^{* A}V_{id}^{B}$, where $V_L$ and $V_R$ are the Cabibbo-Kobayashi-Maskawa
(CKM) matrix and its right-handed analogue.  A crucial new ingredient in $V_R$ is the presence of
(five) additional phases, besides the Dirac one. These can be parametrized as ($U=u,c,t$, $D=d,s,b$)
\be
V^R_{UD}={\rm e}^{i\theta_U}\hat V^R_{UD}{\rm e}^{i\theta_D}\,,
\label{eq:phases}
\ee 
with $\hat V_R$ the mixing matrix in standard CKM form.

The terms in the expressions (\ref{eq:C2AB}), (\ref{eq:C4C9}), (\ref{eq:Cg12}) for the coefficients
should be understood as generated at the decoupling of the relevant heavy states, and thus at
different scales, namely: $M_{W_L}$ or $m_t$ for the current-current and top-dominated loops, $m_c$
for the charm dominated loops etc, and $m_{W_R}$ for the RR current-current.

A similar set of operators $Q_{1,2c}^{AB}$ with the $c$-quark replacing $u$, is also generated by
the short distance physics, and also by renormalization at scales larger than $m_c$.  On the other
hand, the further operators involving the $t$ quark are not explicitly required: for the LL and RR
operators this is due to the GIM cancelation above $m_t$ (also in the running); for the LR ones,
they are only generated at electroweak scale through the LR-mixing $\zeta$.\footnote{Clearly, if
  $Q^{LR}_{1,2t}$ were generated at high scale, they should be taken into account, because due to
  the mixed chirality a GIM cancelation is not effective in the running (see also comments in
  appendix~\ref{app:running}).  Also, some more operators of the form $(\bar sd)_L(\bar dd+\bar
  ss)_{L,R}$ mixing with the penguins are present in general, see~\cite{Buras-Misiak}, but they are
  not generated in the LR model and it is also difficult to generate them in models where new
  physics sets in at scales higher than the electroweak scale.}  Lastly, there are also penguin
operators built through the LR-mixing, which are chirally suppressed and give subleading (negligible)
contributions.

\medskip

From (\ref{eq:Cg12})\ it can be seen that the coefficients $C_{g}^{L,R}$ receive a large
contribution in the LR model, due to the different GIM mechanism.  In fact, the mass insertion on
the external fermion legs in the SM ($m_s$) is replaced in the LR model by a mass insertion inside
the loop ($m_i$).  The loop is then dominated by $m_c$ leading to an enhancement of $m_c/m_s\sim
100$. In addition, the factor $\lambda_c^{AB}/\lambda_{t}^{LL}$ gives a further large enhancement of
$10^3$, which compensates the LR-scale suppression $\zeta$. Both $Q_{g}^L$ and $Q_{g}^R$ are
present, and the LR contribution ends up being a factor $\sim 200$ larger than the SM one, at short
distance:
\begin{equation}
  \frac{|H_{g}^{(LR)}|}{|H_{g}^{(SM)}|}\simeq \frac{2\, m_c F^{LR}_2(x_c)|V_{cd}^* V_{cs}\zeta|}{m_s F_{2}^{LL}(x_t) |V_{td}^* V_{ts}|}\simeq2\times10^5\,\zeta \simeq 200\,.
\end{equation}
Here, the factor 2 accounts for the contributions $LR+RL$, and we considered $\hat V_R\simeq V_L$,
which is a general prediction of minimal LR models~\cite{noi}.

The new phases (\ref{eq:phases}) containted in $V_R$, together with the enhancement above, can
directly induce a sizeable CP violation.  It is therefore important to address the effect of this
operator on $\e'$, which we study along the lines of~\cite{eeg, gabrielli}.  In order to deal with
this low energy phenomenon, two steps are necessary: the first is to renormalize the coefficients at
low energy, in the range of chiral perturbation theory; the second is to use the matrix elements
$\langle2\pi|Q_i|K_0\rangle$, or equivalently to match with chiral perturbation theory.  In the
following section we renormalize the full set of coefficients down to the scale of 0.8\,\GeV, and in
the next we match with the chiral lagrangian in the context of the Chiral Quark Model.

The need to evaluate the Wilson coefficients at such a low QCD scale is dictated by the requirement
to use the matrix elements calculated in the context of Chiral Quark Model in Chiral Perturbation
Theory, whose cut-off is the chiral symmetry breaking scale. In order to assess quantitatively the
scale dependence of the result, we remark that by varying the matching scale between 0.8 and 1 GeV
the chromomagnetic $C_g^{L,R}$ and the current-current ones $C_{1,2}^{LR,RL}$ vary at most by 5\%
and 10\%, respectively.. These uncertainties are well below those of the matrix elements.

\NEG\NEG
\section{Running to low scale}
\label{sec:running}
\NEG

\noindent
The mixing of operators~(\ref{eq:operators}) is described in detail in appendix~\ref{app:running}.
At leading order, the operators can be split into two sets, of opposite chiralities, corresponding
to the two columns in~(\ref{eq:operators}). The low energy coefficients together with the matrix
elements of all the operators are also sufficient to give an estimate of the impact on $\e'$ for
quite a large class of models of new physics.  This will be presented in section~\ref{sec:results}.

In the particular case of the LR model, the low energy coefficients are shown in
table~\ref{tab:runresult2}.  The running takes into account the whole set of operators including the
SM penguins, but we show the result only for the operators containing the LR scale $\beta$ or
$\zeta$ which have an impact on $\e'$, and the magnetic operator $Q_{\g}^{L,R}$ which is also
enhanced. The results are normalized to $\l_u$, to compare with existing calculations. The
coefficients $C_{g}^{L,R}$, compared with the complex part of the SM result, $C_{g}^L(SM)\simeq
0.34\lambda_t$~\cite{eeg}, confirm the important role of $Q_{g}$ from new physics.

\begin{table}[t]
\centering
\tabcolsep=1ex
\begin{tabular}{|l|l|}
\hline
$C_1^{RL,LR}$ & $\lambda_u \,  (1.07)\,|\zeta|e^{\pm i (\a-\theta_{s,d}-\theta_u)}$ \\[.7ex]
\hline
$C_2^{RL,LR}$&  $\lambda_u \,\, (0.80)\, \,|\zeta|e^{\pm i (\a-\theta_{s,d}-\theta_u)}$\\[.7ex]
\hline
$C_1^{RR}$ &   $\lambda_u \, (-0.54) \,\beta e^{-i (\theta_s-\theta_d)}$  \\[.7ex]
\hline
$C_2^{RR}$  &   $\lambda_u \,\, (\,\,1.24)\, \,\beta e^{-i (\theta_s-\theta_d)}$  \\[.7ex]
\hline
$C_{g}^{L,R}$  &   $\lambda_u \, (-10.7) \,\,|\zeta|e^{\pm i (\a-\theta_{d,s}-\theta_c)}$  \\[.7ex]
\hline
$C_{\g}^{L,R}$  &   $\lambda_u \, (-3.31) \,|\zeta|e^{\pm i (\a-\theta_{d,s}-\theta_c)}$  \\[.7ex]
\hline
\end{tabular}
\caption{Coefficients for the dominant new operators in the minimal LR model, evaluated at $\mu = 0.8\,\GeV$.\label{tab:runresult2}}
\end{table}

In the detail of the running it is worth noting that, despite the reduction of $\sim$0.5 due to
their own anomalous dimension, $C_{g}^{L,R}$ receive contributions from $C_{1,2}^{LL,LR,RL,RR}$. The
largest additional contribution is due to $C_{1,2c}^{RL,LR}$ at scales above $m_c$, while the
contributions from $C_{1,2u}^{RL,LR}$ are suppressed by the $u$ quark mass. This is due to the
internal mass insertion in the (two) loop graphs responsible for the operator mixing in the
anomalous dimension matrix, and is an other consequence of the nonchiral nature of these operators.
As a side result, this additional contribution preserves the same combination of phases appearing in
the original short-distance $C_{g}^{L,R}$.

From table~\ref{tab:runresult2}, we can calculate the contributions to $\e'$ of these operators.
This requires the evaluation of matrix elements which we review now for $Q_{g}^{L,R}$.

\NEG\NEG
\section{Bosonization of $Q_{g}$}
\label{sec:bosonization}
\NEG

\noindent
The bosonization of $Q_{g}$ was addressed in~\cite{eeg} in the context of the Chiral Quark Model.
Here we review the computation, which leads a minor numerical correction.

Under chiral $SU(3)_L\times SU(3)_R$ rotations the $Q_{g}$ operators transform as $({\bf 3}_L,{\bf
  3}_R)$, and thus they give rise to particular terms in the chiral Lagrangian.  While by symmetry
arguments there are diverse possibilities (see e.g.~\cite{He:1999xa} in naive dimensional analysis)
in the context of the Chiral Quark Model, only one form arises~\cite{eeg}. This is true in the SM as
with the separate operators $Q_{g}^L$, $Q_{g}^R$. One has
\be
\label{eq:chiralQg}
{\mathcal L_{Q_{g}}}={\rm Tr}\left[\left(\Sigma^\dag X \lambda_- + \lambda_- X^\dag \Sigma\right)D_\mu \Sigma^\dag D^\mu\Sigma\right],
\ee
\begin{figure}[t]
\centerline{\includegraphics[width=\columnwidth]{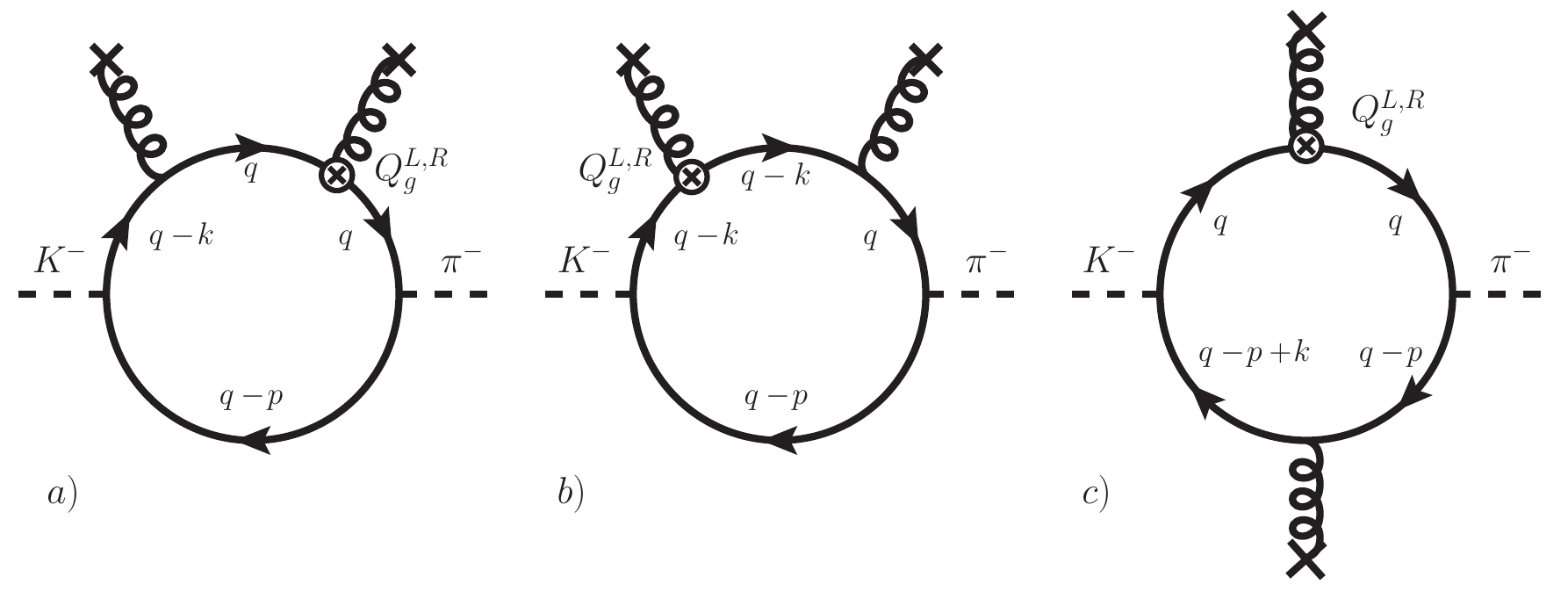}}%
\caption{Diagrams for the bosonization of $Q_{g}$, in the fixed point gauge. Note the
  flow of gluon momentum.}
\vspace*{-1ex}\label{fig:Qg}
\end{figure}%
where $\lambda_-=(\lambda_6-i\lambda_7)/2$, and where the matrix of two coefficients $X={\rm
  diag}(0,G_8^R,G_8^L)$ replaces the single coefficient $G_8^{(4)}$ and the running quark mass
matrix ${\mathcal M}={\rm diag}(m_u,m_d,m_s)$ of the analogous calculation in the SM.  In fact, the
coefficients $C_{g}^{L,R}$ of the $\Delta S=1$ transition induce a tiny breaking of the chiral
symmetry which plays the same role as the $s$ and $d$ quark masses in the SM.  We also observe that
from the point of view of chiral perturbation theory the above lagrangian is $O(p^4)$ in the LL and
RR terms proportional to light quark masses, but $O(p^2)$ in the LR and RL terms, proportional to
$m_c$.  In any case, the coefficients $G_8^{L,R}$ will respectively be proportional to
$C_{g}^{L,R}$. To determine them, it is convenient to evaluate some amplitude both through the above
chiral lagrangian and in the Chiral Quark Model, and compare the results~\cite{eeg}.  We do this in
the `unrotated' picture~\cite{Pich:2004ee}. The simplest process is the off-shell $K^-\to\pi^-$
transition, which at one loop is given by the three diagrams shown in figure~\ref{fig:Qg}. The two
external gluon lines are attached to the gluons in the external thermal QCD vacuum and lead to a
coefficient proportional to the gluon condensate. To deal with the thermal and color average, it is
best to adopt Fock-Schwinger fixed-point gauge ($x^\mu A_\mu = 0$)~\cite{novikov}.  Due to this
gauge choice translations are broken and two fixed `sink' points for the gluon momentum are defined,
chosen here to be $x=0$ and $x$ at the $K$ and $\pi$ vertices.  Then the three diagrams in
figure~\ref{fig:Qg} are
%
\def\Kus{}
\def\piud{}
\def\GSD{}
\newcommand\q{q\!\!\!/}
\renewcommand\k{k\!\!\!/}
\newcommand\p{p\!\!\!/}
\def\Aslash{A\!\!\!/}
\def\FAD{\Delta}
$$
A(K^-\to\pi^-) = \frac{2m^2}{f^2}\int\frac{{\rm d}^4q}{16\pi^4} (a+b+c)
$$
\vspace*{-3ex}
\bea
a=&&\tr[\Kus(\q - \k + m)\Aslash(\q + m)
   q_{g} (\q + m)\piud(\q - \p +
     m)]\notag\\
&& \FAD[{q, m}]^2 \FAD[{q - k, m}] \FAD[{q - p, m}]
\notag\\[1ex]
b=&&\tr[\Kus(\q - \k + m)
   q_{g}(\q - \k + m)\Aslash(\q +
      m)\piud(\q - \p + m)] \notag\\
&&\FAD[{q, m}] \FAD[{q - k,
    m}]^2 \FAD[{q - p, m}]
\notag\\[1ex]
c=&&\tr[\Kus(\q + m)
   q_{g}(\q + m)\piud(\q - \p +
      m)\Aslash(\q + \k - \p + m)] \notag\\
&& \FAD[{q, m}]^2 \FAD[{q - p,
    m}] \FAD[{q - p+ k, m}],
\eea
where $p$ is the $K$ and $\pi$ momentum; $k$ is the incoming gluon momentum;
$\Delta[p,m]=1/(p^2-m^2)$;
$q_{g}=-iG_Fg_sm_s(C_{g}^LL+C_{g}^RR)G^{\alpha\beta}\sigma_{\a\b}/\sqrt{2}$ and $m$ is the
constituent quark mass.  Also (see~\cite{novikov}) $\Aslash=(g_s/2) \gamma_\mu
G_{\mu\nu}\partial/\partial k_\nu$ where a derivative with respect to the gluon momentum has to be
taken, after which $k$ is set to zero.\footnote{Due to the absence of translational invariance, the
  use of the one-external-gluon effective quark propagator in the fixed point
  gauge~\cite{Reinders:1984sr} is not correct in diagram (b), where the gluon momentum flows to the
  origin ($K$) passing through the operator insertion.  This leads to a mismatch between diagrams
  $a$ and $b$ and to a numerical correction of the result in ref.~\cite{eeg}.}  By the same
prescriptions one shows that no external gluon momentum flows through the chromomagnetic
operator. Finally, the two gluon field-strengths are averaged in the gluon condensate,
$G_{\alpha,\beta} G_{\gamma,\delta} \to (\pi ^2/6g_s^2) \left\langle \frac{\alpha }{\pi
  }GG\right\rangle (g_{\alpha \gamma } g_{\beta \delta }-g_{\alpha \delta } g_{\beta \gamma })$.

\begin{table*}[t]
\renewcommand\arraystretch{1.3}
\footnotesize
$
\begin{array}{|l|ccccccccccccccccl|}
\hline
\ \ \ w_i \times 10 & w_{1c}^{RL}&w_{2c}^{RL}&w_{1}^{RL}&w_{2}^{RL}&w_{1c}^{LL}&w_{2c}^{LL}&w_{1}^{LL}&w_{2}^{LL}&w_3&w_4&w_5&w_6&w_7&w_8&w_9&w_{10}&w_{g}^L\\[.5ex]
\hline
\m=m_{W} &0.52&-0.068&160.&52.&0.086&-0.55&2.0&-0.24&-0.14&-3.4&8.1&22.&110.&340.&3.2&0.5&0.0020\\
\m=m_c  &0.&0.&51.&15.&0.&0.&1.9&1.0&0.16&-0.74&2.7&8.0&32.&110.&2.8&1.9&0.0040\\
\m=0.8\,\GeV &0.&0.&42.&11.&0.&0.&2.0&1.2&0.22&-0.55&2.2&6.7&25.&88.&2.8&2.1&0.0043\\
\hline
\end{array}$
\caption{Weigths $w_i$ (times 10) of the coefficients $C_i$ in the determination of $\e'$, 
  applicable to coefficients from different scales. The weights relative to the opposite chirality 
  operators have opposite sign.\label{tab:result}}
\end{table*}

The loop integration produces a term at order zero in $p$ which is canceled in the leading chiral
lagrangian in agreement with the FKW theorem~\cite{FKW}. 
The second order term gives the desired result
\be
A(K^-\!\to\!\pi^-)=
\frac{iG_F}{\sqrt{2}}m_s \frac{C_{g}^L\!+\!C_{g}^R}{16\pi^2} \left\langle\frac\alpha\pi GG\right\rangle p^2 \frac{ (7 +\!1+\!8 )}{48 f^2 m}\,,
\ee
the three numbers being relative to diagrams $a$, $b$, $c$.  Comparing this amplitude with the one
calculated from the chiral lagrangian~(\ref{eq:chiralQg}), one finally finds the coefficients
\be
\label{eq:G8}
G_8^{L,R}=2 \frac{1}{12m} \left\langle\frac\alpha\pi GG\right\rangle \, \frac{G_F}{\sqrt{2}}\frac{m_sC_{g}^{L,R}}{16\pi^2}\,.
\ee
The factor of $2$ corrects $11/4$ appearing in \cite{eeg}, and leads to a 30\% reduction of the
matrix element. This result is also confirmed by a similar calculation within the `rotated' picture.

Before using this result for the calculation of $\e'$, let us note that the additional contribution
from the \emph{off-shell} chromomagnetic $sdg$ vertex, shown in~\cite{Eeg:2008kf} to be of the same
order as $Q_g$ in the SM, is strongly suppressed in the case of nonchiral interactions (again
because the mass insertion happens inside the W loop). 

Finally, a double insertion of $Q_g$, leads directly to (long-distance) $\Delta S=2$ processes and
can be calculated similarly.  However, the process is doubly loop and $G_F$ suppressed and the
result is negligible for both $\Delta M_K$ and $\e$ (see appendix~\ref{app:QgQg}).

\NEG\NEG
\section{Result for $\e'$}
\label{sec:results}
\NEG

\noindent
The direct CP parameter $\e'$ is defined by
\begin{equation}\label{epsilonprime}
  \e'=\frac{i}{\sqrt{2}}\,\omega\left(\frac{{\rm Im} A_2}{{\rm Re} A_2}-\frac{{\rm Im} A_{0}}{{\rm Re} A_{0}}\right)\frac{q}{p}\,e^{i(\delta_{2}-\delta_{0})},
\end{equation}
where $p$, $q$ are the $K^0$, $\overline{K}^0$ mixing parameters and $\omega\equiv A_{2}/A_{0}\simeq
1/22.2$. The ratio $p/q\simeq1$ with an excellent approximation. The isospin amplitudes $A_{I}$
($I=0,2$) are defined from the $\Delta S=1$ effective Hamiltonian as $\langle
(2\pi)_{I}|(-i)H_{\Delta S=1}|K^{0}\rangle =A_{I}e^{i\delta_{I}}$, where $\delta_{I}$ are the strong
phases of $\pi \pi$ scattering. We calculate the imaginary part of the amplitudes, while for the
real part we take the experimental value: ${\rm Re} A_0=3.33\times 10^{-7}\,\GeV$ and ${\rm Re}
A_2=1.49\times 10^{-8}\,\GeV$.

The amplitudes $A_0$ and $A_2$ for the standard operators are collected in
appendix~\ref{app:amplitudes}.  The ones of $Q_{g}^{L,R}$ are easily calculated from the chiral
lagrangian~(\ref{eq:chiralQg}).  One has the isospin decomposition
\be
\label{eq:A0A2Qg}
A^{Q_{g}^{L+R}}_0=\sqrt{\frac{3}{2}}A^{Q_{g}^{L+R}}_\pm\,,\qquad
A^{Q_{g}^{L+R}}_2=0 ,
\ee
where the amplitude for $K^0\rightarrow \pi^+ \pi^-$ is
%
\bea
\label{eq:ApmQg}
  A_{\pm}^{Q_{g}^{L\!+\!R}}\!\!\!&=&\frac{\sqrt{2}}{f^3}m_\pi^2 (G_8^L-G_8^R)=\nonumber\\
&=&\frac{G_Fm_\pi^2}{6mf^3} \left\langle\frac\alpha\pi GG\right\rangle \frac{m_s(C_{g}^L\!-\!C_{g}^R)}{16\pi^2}\,,
\eea
and equals the $K^0\rightarrow \pi^0 \pi^0$ amplitude.  In the following we use $f=93\,\MeV$, and
for the gluon condensate and constituent quark mass we adopt the central values
$\langle\frac\alpha\pi GG\rangle=(334\,\MeV)^4$, and $m=200\,\MeV$~\cite{eegbertfabbreview},
obtained by consistently fitting in the model the $\Delta I=1/2$ selection rule.


Using the running and the matrix elements, one can generically describe the contributions to $\e'$
of the different operators, as a weighted sum of the coefficients contributing from the desired
scales $\mu_n$:
\be
\label{eq:epsprimegeneral}
|\e'|=
\left |\sum_n \sum_i w_i(\mu_n)\, {\rm Im}\,C_i(\mu_n)\right|\,,
\ee
where, at each scale $\mu_n$, the $C_i(\mu_n)$ are the coefficients and $w_i(\mu_n)$ their weights.
The scales $\mu_n$ can be either taken as the ones where the short distance coefficients are
generated, i.e.\ $m_W$, $m_c$, etc., in which case the $w_i$ account for the running and the matrix
elements, or some low energy scale if one includes the running in the $C_i$.  In
table~\ref{tab:result} we collect the numeric values of $w_i$ computed by taking into account the
complete running from a choice of different scales down to 0.8\,\GeV, together with the required
matrix elements (see appendix~\ref{app:amplitudes}).  

As discussed in appendix~\ref{app:amplitudes}, the hadronic uncertainties present in the $\B$
factors of departure from vacuum saturation approximation can be sizable and may vary from 10 to
50\% for the better known operators, to a factor of order one for $Q_g$ and $Q_{1,2}^{LR}$.  For the
SM operators the values adopted in table~\ref{tab:result} are taken from the Chiral Quark Model
calculation~\cite{eegbertfabbreview}.  For the LR operators $Q_{1,2}^{LR}$, a determination is
missing but an guess can be given by noting their similarity with operators $Q_{7,8}$.

For $Q_g^{L,R}$, while the leading order chiral bosonization~(\ref{eq:chiralQg}) results in a
$m_\pi^2/m_K^2$ suppression (see (\ref{eq:ApmQg})), this may cease to be true in higher orders and
may result in a further enhancement. Together with chiral loops, this is likely to lead to order one
correction coefficients $\B^g_{0,2}$ to be added to eqs.~(\ref{eq:A0A2Qg}).  While we stress the
need for a dedicated assessment of these corrections in the view of new physics, in the present
analysis we conservatively assume $\B_{0,2}^g=1$, keeping in mind that a possible enhancement would
make our bounds below stronger.

Clearly, taking the central values of the hadronic matrix elements is sufficient for the scope of
assessing the relevance of contributions beyond the SM, and, in particular, in view of the leading
role of the chromomagnetic operators shown by the present analysis.  Also, the renormalization
evolution is performed at leading order, as is the determination of the starting conditions in
eqs.~(\ref{eq:C2AB}--\ref{eq:Cg12}). Nevertheless, since the penguins can be considered as NLO
contributions, a NLO correction to the current-current starting conditions was also inserted
(see~\cite{standardoperators2}). In this respect let us again remark that while a NLO analysis is
necessary for the SM, it is not crucial for assessing the constraint from $\e'$ on new physics.  We
believe table~\ref{tab:result} with formula~(\ref{eq:epsprimegeneral}) to be useful in analyzing the
impact of $\e'$ for quite a wide class of new physics models.

\NEG\NEG
\section{Consequences for the LR model}
\label{sec:LR}
\NEG

\noindent
We can finally estimate the numerical impact on $\e'$ of the new physics operators in the case of
LR-symmetry, using the values of the low energy coefficients summarized in
table~\ref{tab:runresult2} and the last line of table~\ref{tab:result} (or equivalently the
appropriate short distance coefficients with the first lines).  We find
\bea
|\epsilon'_{LR}|&\simeq&
\bigg|\ 
|\zeta|\,1.25\big[ \sin(\alpha-\!\theta_u-\!\theta_d)+\sin(\alpha-\!\theta_u-\!\theta_s)\big]\notag\\[1ex]
 && {}+
|\zeta| \,0.0010\big[\sin(\alpha-\!\theta_c-\!\theta_d)+\sin(\alpha-\!\theta_c-\!\theta_s)\big]
\notag\\[1ex]
&&{}+
\beta\,0.013\sin(\theta_d-\!\theta_s)\,\bigg|\,,
\label{epsprimenum}
\eea
where the first line is due to the dominant $Q_{1,2}^{LR,RL}$, the second to $Q_{g}^{L,R}$, and the
last to $Q_{1,2}^{RR}$.\footnote{Also $\lambda_c^{LL}$ is complex, but its phase is O(1/1000) and is
  subleading in this expression.} The penguin contributions are only responsible for minor
corrections in the above numeric result.

For \TeV\ LR-scale ($\beta\sim 10^{-3}$), we see that the above contributions can give overdominant
contributions to $\e'$, even having assumed similar left and right quark mixing angles, as in the
minimal LR models.  This is true both for $Q_{g}^{L,R}$ and for the other operators.  Their impact
may be different and it depends, in addition to the LR scale $\beta$, on the CP phase $\alpha$ and
on the extra phases in $V_R$.  The actual implications for a given LR model depend thus on the
available freedom in choosing these phases. Let us describe the bounds in different scenarios,
assuming that the new physics can contribute as much as 100\% to~$\e'$.

Assuming generic O(1) free phases, the LR scale is constrained to lie above a large limit
$M_{W_R}\gtrsim 25\,$TeV.  In fact this amounts to the highest limit on the right-handed scale.  It
is also worth recalling that for order one phases another large bound of about $M_{W_R}\gtrsim
15$TeV results from $\e$, while limits from $B$ mass difference and CP violation are less
stringent~\cite{moha, noi}.  The argument can however be turned around and~(\ref{epsprimenum})\ can
be used to put constraints on the phases, in the scenario of TeV LR-symmetry. 

In the minimal LR models, phases are either strictly predicted or are free, depending on the choice
of LR-symmetry, which can be generalized parity (\P, exchanging fermions $\psi_L\leftrightarrow
\psi_R$) or generalized charge conjugation (\C, exchanging fermions $\psi_L\leftrightarrow
\psi_R^c$), following the analysis of ref.~\cite{noi}.  

An important common constraint resulting in both cases from $\e$ is that $\theta_d-\theta_s$ is
close to 0 (or marginally to~$\pi$) at least as $10^{-2}$. This implies that the contribution of
$Q^{RR}_{1,2}$ to $\e'$ can be neglected (the last line in~(\ref{epsprimenum})).  This can already
be seen from the low energy coefficients in table~\ref{tab:runresult2}.  From the same
table~\ref{tab:runresult2} it is also important to note an other consequence of the $\e$ constraint,
namely that thanks to $\t_s\simeq \t_d$ the coefficients are practically complex conjugated under
$L\leftrightarrow R$ exchange.  As a result, the contributions to $\e'$, which are proportional to
$L-R$ combinations (as $C_g^L-C_g^R$, see~\ref{eq:ApmQg}) are purely imaginary and thus with maximal
imaginary part.  On the contrary, only the combinations $L+R$ would enter in other CP-violating
observables involving an even number of mesons, making the imaginary part suppressed by
$\t_s-\t_d\lesssim 10^{-2}$.  This is the case for instance for the contribution of the magnetic
operators $O_\gamma^{L,R}$ to the CP asymmetry in $K\to \pi e^+e^-$, which is thus suppressed,
despite the enhancement of the Wilson coefficient.  This situation can be contrasted with the one
occurring e.g.\ in supersymmetric models, where a correlation between $\e'$ and $K\to \pi
\ell^+\ell^-$ can be inferred~\cite{colangelo}. Nevertheless, the enhancement of $C_\gamma^{L,R}$
would survive in CP-asymmetries with an odd number of mesons, like $K\to \pi\pi \ell^+\ell^-$ (whose
analysis brings in the leptonic sector of the LR models and is beyond our study).

Let us then describe the impact of the first two lines in~(\ref{epsprimenum}) to $\e'$, for the two
possible choices of LR-symmetry.

\medskip

In the case of \P, due to the hermiticity of the Yukawa couplings, the phases in $V_R$ are all
predicted in terms of the phase $\alpha$, and they are all close to 0 or $\pi$. The neutron EDM then
poses a strong constraint which together with $\e$ and $\e'$ leads to the strong limit
$M_{W_R}>8\text{--}10\,\TeV$~\cite{moha}.  As discussed in~\cite{noi}, a TeV-scale LR symmetry is
still allowed by resorting to an unappealing fine-tuning with the QCD strong phase $\bar\theta$.  In
this case, the $\e'$ gives alone a bound, because from $\e$ one must have $x \sin \alpha\simeq
10^{-3}$~\cite{moha, noi}, so that the limit $x\to0$ i.e.\ $\zeta\to0$ that would suppress $\e'$ in
(\ref{epsprimenum}) is not permitted. Then, by using the values for the predicted phases (see
eq.~(29) in ref.~\cite{moha}) and exploiting conservatively the free signs, in particular $u$ and
$c$ opposite, one finds the numeric result\footnote{This corresponds to the case
  $|\theta_d-\theta_s|\sim 10^{-2}$. There is also a second possibility, with
  $|\theta_d-\theta_s|\sim \pi+O(10^{-2})$ and $x \sin \alpha\approx 10^{-2}$, but it is disfavored
  by the $B_{d,s}$ mass differences and CP-violation.  Moreover, in this case there are cancelations
  in each line of (\ref{epsprimenum}) and the situation is more ambiguous, since it depends on the
  precision to which the equality $V_L\sim V_R$ of mixing angles holds. We recall that with a global
  numerical study of this model~\cite{noi}, the angles could be deviated as much as 20\%, which
  would spoil these cancelations in $\e'$. Therefore also in this case one can expect a dominant
  contribution as in (\ref{epsprimenumcaseP}).}
\bea
|\epsilon'_{LR}|&\simeq& 5.7 \, 10^{-6} (\beta/10^{-3})\,.
\label{epsprimenumcaseP}
\eea
This result holds in the natural regime $x<0.1$, where analytic expressions for the phases are
available.  Comparing $\e'_{LR}$ with $100\%\,(50\%)$ of $|\e'|_{exp.}\simeq 3.92\times10^{-6}$, we
obtain the constraint $M_{W_R}\gtrsim 3\,(4)\,\TeV$.  Here the main contribution comes from
$Q_{1,2}^{LR,RL}$, while $Q_{g}^{L,R}$ contribute subdominantly, with the effect of softening the
limit with respect to the conclusions of~\cite{noi}.\footnote{We also do not agree with the strong
  bound derived in ref.~\cite{Chen:2008kt}, using only the isospin-2 amplitude of the operator
  $Q_1^{LR}$.}

\medskip

In the more interesting case of \C\ as LR-symmetry, the phases (\ref{eq:phases}) in $V_R$ are free.
This time, the bound from $\e$ together with the $B_{d,s}$ systems put the stronger constraint
$\theta_d-\theta_s< 10^{-3}$~\cite{noi}.  As a result, from (\ref{epsprimenum}) one has
\bea
|\epsilon'_{LR}|&\simeq&
|\zeta|\Big|
2.50\sin(\alpha-\theta_u-\theta_d)\notag\\
&&\quad\qquad\qquad
+0.0020\sin(\alpha-\theta_c-\theta_d)\Big|\,.
\label{epsprimenumcaseC}
\eea
The first line due to $Q_{1,2}^{LR,RL}$ is dominant, but one can note that the phase combination
appearing there is independently constrained by the neutron EDM: as studied
in~\cite{moha,Xu:2009nt,noi} one has $|\zeta|\sin(\alpha-\!\theta_u-\!\theta_d)< 10^{-7}$. This
implies that the current-current contribution can be neglected here, and thus we are left with the
dominance of the second line, due to $Q_{g}^{L,R}$.  For $\e'_{LR}$ to at most saturate the
experimental value one has the constraint
\be
\label{eq:Climit}
|\zeta| |\sin(\alpha-\theta_c-\theta_d)|< 2.0 \times 10^{-3}
\ee
(or a correspondingly more stringent one for a subdominant $|\e'|_{LR}$). This represents a
correlated bound between the phases and the LR-symmetry scale/mixing.  For unconstrained phases
$\theta_c+\theta_d$ one would require $M_{W_R} > 2.8\,\TeV$, or vanishing LR-mixing. 
%

This constraint is similar to the one reported in ref.~\cite{He-Tandean-Valencia},
albeit a different evaluation of the $Q_g$ matrix element was there adopted and only the `charm'
couplings were considered. As discussed above, the uncertainty (possible enhancement) in the matrix
element of $Q_g$ can strenghten this bound by an order one factor.

The bound~(\ref{eq:Climit}) is also analogous to the limit inferred from the $s\to
d\gamma$ decays~\cite{He:1999ik,He-Tandean-Valencia}, which through $Q_\gamma^{L,R}$ involve the
same enhancement and the same phases as the chromomagnetic operator.

\NEG
\section{Summary and conclusions}
\label{sec:conclusions}
\NEG

\no In this work we addressed the effect on $\e'$ of new physics in the chromomagnetic dipole
operators, which can have a huge enhancement with respect to the SM, especially in the presence of
nonchiral interactions.  The paradigmatic example for this effect appears in minimal Left-Right
symmetric theories, where the $W_L$-$W_R$ gauge-boson mixing leads to an enhancement of $10^5$ in
the short distance loop coefficient, so that even with a scale of new physics in the TeV region, an
enhancement of two orders of magnitude results. Together with the presence of new phases in the
Right quark mixing matrix, this can lead to a dramatic impact in $\e'$.

To evaluate quantitatively the effect, we considered the dipole operators together with the full set
of four quark operators which can give rise to CP violation in $K\to \pi\pi$ decays.  We considered
their renormalization and mixing (at leading order) from short distance to the low scale of matching
with chiral perturbation theory, where the matrix elements can be estimated.  For the chromomagnetic
dipole operators we reevaluated the corresponding matrix element in the context of the Chiral Quark
Model (correcting the previous existing calculation).  Tor the SM operators we adopted the estimates
consistently determined in previous analysis of $\e'/\e$~\cite{eegbertfabbreview}. These were also
used for an estimate for the LR current-current operators $Q_{1,2}^{LR}$ (see discussion in
section~\ref{sec:results} and appendix~\ref{app:amplitudes}).  The set of high energy operators
considered is fairly complete, and can serve also for estimates of the impact of other new physics
models on $\e'$.

We applied the results to the case of the minimal Left-Right model, showing that $\e'$ receives
contributions from the chromomagnetic operator as well as from the current-current ones
$Q_{1,2}^{LR,RL}$.  These are in general large, but we noted that they are severely constrained by
the nEDM, with the result that the chromomagnetic operators turn out to be dominant.  One can expect
this to be a fairly generic situation, because new CP phases contributing to $Q_{1,2}^{LR,RL}$ are
usually contributing as well to the nEDM.

In the LR model, focusing on the case of generalized charge-conjugation $\C$ taken as LR-symmetry,
where new phases are free, this allows us to derive a correlated constraint between the LR-gauge
boson mixing (or LR-symmetry scale) and the relevant phases.
The bound for arbitrary phases amounts to $\zeta\lesssim 10^{-3}$ (or equivalently $M_{W_R}>2.8\,\TeV$).

The potential dominance of the $Q_{g}$ contribution can then lead to constraints on processes involving the
same combination of phases. This is true for instance for the magnetic operators $O_\gamma^{L,R}$
which are also GIM-enhanced, and as we argued will enter in $K\to \pi\pi e^+e^-$ CP-violating
asymmetries. Also, $(\alpha-\theta_c-\theta_d)$ enters in the charmed mesons physics the analysis of
which is beyond the scope of this work. Nevertheless let us point out that it enters the decays of
the $D$ meson via $c\to u\gamma$, whose short distance contribution is overwhelmed by the long
distance ones~\cite{Buchalla:2008jp}, but also it enters the CP-violation in the $D\to KK,\pi\pi$
channels, for which anomalous signals have been reported by the LHCb collaboration~\cite{LHCb}. The
interesting analysis of the related charm physics in the LR-models will be the subject of a separate
work.

\NEG\NEG
\section*{Acknowledgements}
\NEG

\no 
We thank Diptimoy Ghosh for calling our attention on Eq. (12) that was initially misreported. 
We thank G. Senjanovi\'c for useful comments on the manuscript and M. Nemev\v sek for
discussions.  SB is associated to the theoretical particle physics group at SISSA, and partially
supported by MIUR and the EU UNILHC-grant agreement PITN-GA-2009-237920.  JOE is partially supported
by the norwegian research council.  FN would like to thank ICTP for hospitality during the
development of this work.


\appendix

\begin{widetext}

\baselineskip=1.05\baselineskip

\section{Loop functions}
\label{app:loop}

\noindent
The loop functions relevant for the SM and the LR model are~\cite{Misiak, Ecker:1985vv, lim,buras1998}.
\bea
&&F_1^{LL}=\frac{x_j(-18+11 x_j+x_j^2)}{12 (x_j-1)^3}-\frac{(4-16 x_j+9 x_j^2) \ln x_j}{6 (x_j-1)^4}\,,\qquad
E_1^{LL}=-\frac{x_j^2(5x_j^2-2x_j-6)}{18(x_j-1)^4}\ln x_j+\frac{19x_j^3-25x_j^2}{36(x_j-1)^3}+\frac{4}{9}\ln x_j\notag\\[1ex]
&&F_2^{LL}=\frac{x_j(2+3x_j-6x_j^2+x_j^3+6x_j \ln x_j)}{4 (x_j-1)^4}\,,\qquad 
\qquad \qquad \quad E_2^{LL}=\frac{x_j (8x_j^2+5x_j-7)}{12(x_j-1)^3}+\frac{x_j^2(2-3 x_j)}{2(x_j-1)^4}\ln x_j\notag\\[1ex]
&&F_2^{LR}=\frac{-4+3x_j+x_j^3-6x_j \ln x_j}{2(x_j-1)^3}\,,\qquad
\qquad \qquad \qquad \qquad
E_2^{LR}=\frac{5x_j^2-31x_j+20}{6(x_j-1)^2}-\frac{x_j(2-3 x_j)}{(x_j-1)^3}\ln x_j.
\eea
where $x_j=(\frac{m_j}{M_W})^2$, $j=u,c,t$.  In addition to these, one has
$F^{RR}_{1,2}=F^{LL}_{1,2}(\beta x_i)$ and similarly for $E_2^{RR}$.

\section{Running of all $\Delta S=1$ operators}
\label{app:running}

\noindent 
For our purposes, the relevant operators are the $Q_i$ appearing in (\ref{eq:operators}) plus the
eight $Q_{1,2\,c}^{AB}$ where $u$ quark is replaced by $c$.  At leading order (LO) the $LR$
operators mix only with $Q_{g,\g}^L$ (in addition to themselves). Similarly, the $RL$ ones mix only
with $Q_{g,\g}^R$.  The operators can thus be divided in two decoupled sets of 18 operators each,
related by the exchange $L\leftrightarrow R$:
\bea
&&\{Q_{1c}^{RL},Q_{2c}^{RL},Q_{1}^{RL},Q_{2}^{RL},Q_{1c}^{LL},Q_{2c}^{LL},Q_{1}^{LL},Q_{2}^{LL},Q_3,Q_4,Q_5,Q_6,Q_7,Q_8,Q_9,Q_{10},Q_{g}^L,Q_{\g}^L\}\,,\\[.7ex]
&&\{Q_{1c}^{LR},Q_{2c}^{LR},Q_{1}^{LR},Q_{2}^{LR},Q_{1c}^{RR},Q_{2c}^{RR},Q_{1}^{RR},Q_{2}^{RR},Q_3',Q_4',Q_5',Q_6',Q_7',Q_8',Q_9',Q_{10}',Q_{g}^R,Q_{\g}^R\}\,,
\eea
as in the two columns of eq.~(\ref{eq:operators}).  The corresponding vectors of coefficients in the
two sets, $\vec C_{L,R}(\mu)$, evolve separately according to the renormalization group equation
\be
\left(\frac\partial{\partial \ln\mu}+\beta(g)\frac\partial{\partial g}+\gamma_{m_i} \frac\partial{\partial \ln m_i}\right) \vec C_{L,R}(\mu)=\gamma^T(\mu)\vec C_{L,R}(\mu)\,,
\ee
with $i=u,s,c$.  The $18\times18$ anomalous dimension matrix $\gamma$ is the same in the L and R
sectors and reads~\cite{Misiak, ciuchini}
\begin{equation*}
\small
\frac{\alpha_s}{4\pi}\left(
\begin{array}{ccccccccccc}
 -16 & 0 & 0 & 0 & 0 & 0 & 0 & 0 & 0 & 0 & 0 \\
 -6 & 2 & 0 & 0 & 0 & 0 & 0 & 0 & 0 & 0 & 0 \\
 0 & 0 & -16 & 0 & 0 & 0 & 0 & 0 & 0 & 0 & 0 \\
 0 & 0 & -6 & 2 & 0 & 0 & 0 & 0 & 0 & 0 & 0 \\
 0 & 0 & 0 & 0 & -2 & 6 & 0 & 0 & 0 & 0 & 0 \\
 0 & 0 & 0 & 0 & 6 & -2 & 0 & 0 & -2 & \frac{2}{3} & -2 \\
 0 & 0 & 0 & 0 & 0 & 0 & -2 & 6 & 0 & 0 & 0 \\
 0 & 0 & 0 & 0 & 0 & 0 & 6 & -2 & -2 & \frac{2}{3} & -2 \\
0 & 0 & 0 & 0 & 0 & 0 & 0 & 0 & -\frac{22}{9} & \frac{22}{3} & -4 \\
 0 & 0 & 0 & 0 & 0 & 0 & 0 & 0 & 6-2 n_f & -2+\frac{2 n_f}{3} & -2 n_f \\
 0 & 0 & 0 & 0 & 0 & 0 & 0 & 0 & 0 & 0 & 2 \\
 0 & 0 & 0 & 0 & 0 & 0 & 0 & 0 & -\frac{2 n_f}{9} & \frac{2 n_f}{3} & -\frac{2 n_f}{9} \\
 0 & 0 & 0 & 0 & 0 & 0 & 0 & 0 & 0 & 0 & 0 \\
 0 & 0 & 0 & 0 & 0 & 0 & 0 & 0 & \frac{1}{9} (n_f-3 n_u) & \frac{1}{3} (3 n_u-n_f) & \frac{1}{9} (n_f-3 n_u) \\
 0 & 0 & 0 & 0 & 0 & 0 & 0 & 0 & \frac{2}{9} & -\frac{2}{3} & \frac{2}{9} \\
 0 & 0 & 0 & 0 & 0 & 0 & 0 & 0 & \frac{1}{9} (n_f-3 n_u) & \frac{1}{3} (3 n_u-n_f) & \frac{1}{9} (n_f-3 n_u) \\
 0 & 0 & 0 & 0 & 0 & 0 & 0 & 0 & 0 & 0 & 0 \\
 0 & 0 & 0 & 0 & 0 & 0 & 0 & 0 & 0 & 0 & 0
\end{array}
\right.\qquad\qquad
\end{equation*}
\vspace*{-1.5ex}
\begin{equation}
\small
\qquad\qquad\left.
\begin{array}{ccccccc}
0 & 0 & 0 & 0 & 0 & -8 \frac{m_c}{m_s} & \frac{320}{3}\frac{m_c}{m_s} \\
 0 & 0 & 0 & 0 & 0 & \frac{16}{3}\frac{m_c}{m_s} & \frac{128}{9}\frac{m_c}{m_s} \\
 0 & 0 & 0 & 0 & 0 & -8 \frac{m_u}{m_s} & \frac{320}{3}\frac{m_u}{m_s} \\
 0 & 0 & 0 & 0 & 0 & \frac{16}{3}\frac{m_u}{m_s} & \frac{128}{9}\frac{m_u}{m_s} \\
 0 & 0 & 0 & 0 & 0 & 6 & 0 \\
 \frac{2}{3} & 0 & 0 & 0 & 0 & \frac{140}{27} & \frac{832}{81} \\
 0 & 0 & 0 & 0 & 0 & 6 & 0 \\
 \frac{2}{3} & 0 & 0 & 0 & 0 & \frac{140}{27} & \frac{832}{81} \\
 \frac{4}{3} & 0 & 0 & 0 & 0 & (\frac{280}{27}+6 n_f) & -\frac{928}{81} \\
 \frac{2 n_f}{3} & 0 & 0 & 0 & 0 & \left(12+\frac{140 n_f}{27}\right) & -\frac{4}{9} \left(\frac{8 n_f}{9}+12 (n_f-3 n_u)\right) \\
 -6 & 0 & 0 & 0 & 0 & (-\frac{28}{3}-6 n_f) & \frac{64}{9} \\
 -16+\frac{2 n_f}{3} & 0 & 0 & 0 & 0 & \left(-8-\frac{238 n_f}{27}\right) & -\frac{4}{9} \left(\frac{8 n_f}{9}-12 (n_f-3 n_u)\right) \\
 0 & 2 & -6 & 0 & 0 & 0 & 0 \\
 \frac{1}{3} (3 n_u-n_f) & 0 & -16 & 0 & 0 & 0 & 0 \\
 -\frac{2}{3} & 0 & 0 & -2 & 6 & 0 & 0 \\
 \frac{1}{3} (3 n_u-n_f) & 0 & 0 & 6 & -2 & 0 & 0 \\
 0 & 0 & 0 & 0 & 0 & \frac{28}{3} & -\frac{32}{9} \\
 0 & 0 & 0 & 0 & 0 & 0 & \frac{32}{3}
\end{array}
\right)
\end{equation}%
where $n_f$ and $n_u$ are respectively the number of active quarks, and of active up-type
quarks. The mixing of the $Q_{1\,\ldots,10}$ operators with the (chromo)magnetic operators appears at
NLO~\cite{ciuchini}, and we adopt the anomalous dimensions in the HV scheme.
Note, 
the off-diagonal terms in the last two columns carry explicitly the ratio of mass insertions
responsible of the operator mixing.  In fact, while for the mixing of the $LL$ and penguin operators
with the magnetic ones the mass insertion on the external legs is $m_s$ and coincides with the
normalization of $Q_{g,\gamma}$, for the mixing $Q_{1,2}^{LR,RL}\to Q_{g,\g}$ the mass insertion is
that of the internal quark ($m_u$ or $m_c$), breaking the usual GIM cancelations. It also follows
that the traditional description in terms of the $y_i$ and $z_i$ variables~\cite{standardoperators2}
is no longer appropriate, and we need to perform the running of the whole vector of coefficients
$\vec C_{L}$ (and $\vec C_{R}$). Clearly, for the SM part the results coincide, as the GIM
cancelation is effective in the mixing with the dipole operators. Finally, the operators involving
the charm quark are integrated out at their threshold, and accordingly the anomalous dimension
matrix is projected, below this scale, on the remaining set of low energy operators
$Q_{1,2}^{AB},Q_{3\text{--}10},Q_g^L,Q_{\gamma}^L$.

We perform the running choosing $\a_s(M_Z)=0.1176$, and with starting coefficients introduced
separately at the relative scales of decoupling. The running is performed down to 0.8\,\GeV\ where
the matrix elements are evaluated.

\section{Amplitudes}
\label{app:amplitudes}

For all the operators the amplitudes $A_{0,2}$ are expressed in terms of their $K\to(\pi\pi)_{I=0,2}$
matrix elements $\langle Q_i\rangle_{0,2}$:
\begin{equation}
A_0= \sum_i C_i \langle Q_i\rangle_0\,,\qquad
A_2 = \sum_iC_i \langle Q_i\rangle_2\,.
\end{equation}



We report here the matrix elements for the relevant $Q_{1,2}^{LL,LR,RL,RR}$~\cite{eegbertfabbreview,Ecker:1985vv}:
%

\begin{eqnarray}\label{elBert}
\langle Q_1^{LL}\rangle_0 =-\langle Q_1^{RR}\rangle_0 =-\frac{1}{3\sqrt{6}}X \, \B_0^{1}\,,\qquad&\qquad&
\langle Q_1^{LL}\rangle_2 =-\langle Q_1^{RR}\rangle_2 =\frac{4}{3\sqrt{3}}X \, \B_2^{1}\,,\nonumber\\[2ex]
\langle Q_2^{LL}\rangle_0 =-\langle Q_2^{RR}\rangle_2=\phantom{-}\frac{5}{3\sqrt{6}}X \,\B_0^{2}\,,\qquad&&
\langle Q_2^{LL}\rangle_2 =-\langle Q_2^{RR}\rangle_2=\frac{4}{3\sqrt{3}}X \, \B_2^{2}
\end{eqnarray}
\begin{eqnarray}
\langle Q_1^{LR}\rangle_0 =-\langle Q_1^{RL}\rangle_0=\frac{\sqrt{2} (X+9 Y+3 Z)}{3 \sqrt{3}}\,\B_0^{1,LR}\,,
\qquad&&
\langle Q_1^{LR}\rangle_2 =-\langle Q_1^{RL}\rangle_2=\frac{1}{3} \sqrt{\frac{1}{3}} (X-6 Z) \, \B_2^{1,LR}\,,\nonumber\\[2ex]
\langle Q_2^{LR}\rangle_0 =-\langle Q_2^{RL}\rangle_0=\frac{\sqrt{2} (3 X+3 Y+Z)}{3 \sqrt{3}}\,\B_0^{2,LR}\,,
\qquad&&
\langle Q_2^{LR}\rangle_2 =-\langle Q_2^{RL}\rangle_2=\frac{1}{3} \sqrt{\frac{1}{3}} (3 X-2 Z) \, \B_2^{2,LR}\,,
\label{eq:QLR02}
\end{eqnarray}
with
\begin{align}
X & \equiv -\langle \pi^{-}|\bar{d}\gamma_{\mu}\gamma_{5}u|0\rangle\langle\pi^{+}|\bar{u}\gamma^{\mu}s|\overline{K}^{0}\rangle=i\sqrt{2}f_{\pi}(m_{K}^{2}-m_{\pi}^{2})\simeq 0.03i\, \GeV^{3} \notag \\[.7ex]
Y & \equiv -\langle\pi^{+}\pi^{-}|\bar{u}u|0\rangle\langle 0|\bar{d}\gamma_{5}s|\overline{K}^{0}\rangle=i\sqrt{2}f_{K}A^2\simeq0.22i\, \GeV^3 \notag \\[.7ex]
Z & \equiv -\langle \pi^{-}|\bar{d}\gamma_{5}u|0\rangle\langle\pi^{+}|\bar{u}s|\overline{K}^{0}\rangle=i\sqrt{2}f_{\pi}A^2\simeq0.18i\, \GeV^{3}\,,
\end{align}
where $A\equiv m_{K}^{2}/(m_{s}+m_{d})$, and $f_{\pi,K}$ the $\pi$ and $K$ decay constants, and the
quark masses are evaluated at $\m=0.8\,\GeV$ (i.e.\ $m_s\simeq 200\,\MeV$).  Since we use the matrix
elements at $\m=0.8\,\GeV$, also the $\B_i$ coefficients of departure from vacuum saturation have to
be evaluated at this scale.

\medskip
 
The SM ones, determined in the Chiral Quark Model via a phenomenological approach based on the fit
of the $\Delta I=1/2$ rule in $K\to\pi\pi$ decays, can be taken from ref.~\cite{eegbertfabbreview}
(see table~VI), where one can also find the ``correlated'' matrix elements for the operators $Q_{3,
  \ldots, 10}$. For the above current-current operators one finds the central values
${\B^{1}_0}\simeq 9.5$, ${\B^{2}_0}\simeq 2.9$, ${\B^{1,2}_2}\simeq 0.41$. For the gluonic and
electromagnetic penguins relevant to $\e'/\e$ it is found $B^{6}_{0} \simeq 1.6$ and $B^{8}_2
\simeq 0.92$.

\medskip

Concerning the $\B^{1,2,LR}$, their evaluation is still lacking both in the Chiral Quark Model as on
the lattice, and to our knowledge also in $1/N_c$ expansion.  Some hints can be derived from the
observation that the electromagnetic penguins $Q_{7,8}$ transform as $({\textbf8}_L,{\textbf8}_R)$ as do the
$Q_{1,2}^{LR}$. Then, their leading bosonization and chiral loops coincide (see~\cite{antonelli}) so
that one can expect the $\B$ parameters of $Q_{1,2}^{LR}$ to be very similar to those of $Q_{8,7}$.
For the isospin-2 amplitudes this correspondence has even been argued to be exact~\cite{Chen:2008kt}
so that using the results reported in~\cite{eegbertfabbreview} we can set
$\B_2^{1,2,LR}=\B^{7,8}_2\simeq 0.92$.  For isospin-0 amplitudes, the larger $\B_0^{7,8}\simeq 2.5$
hint for $\B_0^{1,2,LR}$ also larger than one.  In general, this is in accordance with the strong
phases from pion rescattering in final state interactions which point to a correction factor of
$\sim1.4$~\cite{Truong:1988zp,Pallante:1999qf}, and also more simply with the correction factor
traditionally applied to $Y$ in vacuum saturation to account for the renormalization to the $K$
scale in the pion matrix element, $(1+m_K^2/\Lambda_{\chi PT}^2)\sim 1.5$, which enhances the
isospin-0 amplitudes in eq.~(\ref{eq:QLR02}).  Therefore for the present analysis we adopt the
conservative choice of central values $B_0^{1,2,LR}\simeq 2$ with $O(1)$ uncertainty.  For the
Left-Right model, the impact of the isospin-0 amplitudes of $Q_{1,2}^{LR}$ is fortunately limited.
The result in the first line of (\ref{epsprimenum})\ changes by 20\% within a 1-3 range of
$B_0^{1,2,LR}$.

\eject

\section{Double insertion of $Q_{g}$ leading to $\Delta S=2$}
\label{app:QgQg}


\hangafter=4\hangindent=-17em 
\noindent
The gluonic dipole operators can give effects also on the $\Delta S=2$ processes, via its double
insertion or via its insertion together with other $\Delta S=1$ operators, most notably the SM
current-current operators $Q_{1,2,}^{LL}$ which have large (real) coefficients. These combinations
constitute true long-distance contributions.  A first estimate was given in~\cite{Donoghue:1985ae}
and the resulting constraint is not relevant for $\Delta M_K$ and is marginal for
$\e$~\cite{He-Tandean-Valencia}.  In the presence of the whole set of new physics operators also
chiral loops with multiple possible insertions should be evaluated, and we leave that for a future
analysis.  However, a simple double insertion of $Q_{g}$ can be readily estimated.  It leads to
$K$-$\bar K$ mixing through the diagram on the right. The two external gluons are averaged in the
vacuum gluon condensate, and the total mixing hamiltionian is:
\be
H_{K\bar K}=-\frac83\frac{M_K^2}{f^2}\frac{G_F^2}2 m_s^2
\frac{\left(C_{g}^L\right)^2+\left(C_{g}^R\right)^2}{16\pi ^2}
\left\langle \frac{\alpha}{\pi }GG\right\rangle   K\bar K
\ee
Considering from table~\ref{tab:runresult2} the low energy values $C_{g}^{L,R}\simeq 2.7 |\zeta|
{\rm e}^{\pm i(\alpha-\theta_c-\theta_{d,s})} $, we find the impact on the $K\bar K$ mixing to
be negligibly small, despite the huge enhancement of the dipole loop:  from  $\Delta M_K
= Re(H_{K\bar K})/2 M_K$ we have
\be
\label{deltaMKQg}
\Delta M_K^{Q_{g}^{L+R}} \lesssim 10^{-21} \,\GeV\,, \qquad \text{(for $\zeta\lesssim 10^{-3}$)}
\ee

\vspace*{-14\baselineskip}
\hfill\vbox to 0pt{\hsize=15em
\includegraphics[width=.24\textwidth]{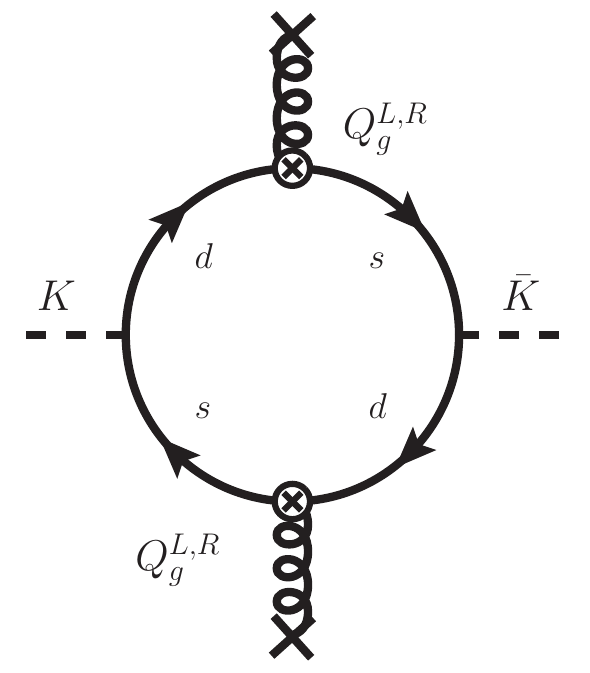}
\vss}
\vspace*{13\baselineskip}

\noindent
which is six orders of magnitude less than the experimental value.  Similarly, also the effect on
$\e$ is negligible: we have $\e\sim 0.3 |\zeta|^2 \cos(2\a) (\theta_d-\theta_s)$. Since
$\zeta<10^{-3}$ and $\t_d-\t_s$ is at most $10^{-2}$, this gives no constraint.


\end{widetext}

\pagebreak[3]

\end{document}